\documentstyle[aps,epsfig]{revtex}
\begin{document}
\begin{titlepage}
\pagestyle{empty}
\baselineskip=18pt
\rightline{NBI--96--48}
\rightline{March, 1996}
\baselineskip=15pt
\vskip .2in
\begin{center}
{\large{\bf Bose-Einstein Correlations from Opaque Sources}}
\end{center}
\vskip .2truecm
\begin{center}
Henning Heiselberg 

{\it NORDITA, Blegdamsvej 17, DK-2100 Copenhagen \O., Denmark}

and Axel P.~Vischer

{\it Niels Bohr Institute, DK-2100, Copenhagen \O, Denmark.}

\end{center}

\vskip 0.1in
\centerline{ {\bf Abstract} }
\baselineskip=15pt
\vskip 0.5truecm
Bose-Einstein correlations in relativistic heavy ion collisions are
very different for opaque sources than for transparent ones. The
Bose-Einstein radius parameters measured in two-particle correlation functions
depend sensitively on the mean free path of the particles. In
particular we find that the outward radius parameter for an opaque source is
{\it smaller} than the sideward radius parameter 
for sufficiently short duration
of emission. A long duration of emission can compensate the
opacity reduction of the longitudinal radius parameter and explain the
experimental measurements of very similar side- and
outward radius parameters.

\end{titlepage} 

\baselineskip=15pt
\textheight8.9in\topmargin-0.0in\oddsidemargin-.0in

\section{Introduction}

Bose-Einstein interference of identical particles (pions, kaons,
etc.), emitted from the collision zone in relativistic heavy ion
collisions shows up in correlation functions and is an important tool
for determining the source at freeze-out. It is commonly
assumed that the source is cylindrical symmetric
around the beam axis and {\it transparent}, meaning that
the detector receives particles from all over the source. The
radius parameter outwards or towards the detector, $R_o$, is found to be
{\it larger} \cite{Heinz,Csorgo} than
the radius parameter, $R_s$, sidewards or perpendicular to the detector
\begin{equation}
   R_o^2 = R_s^2 \;+\; \beta_o^2\delta\tau^2 \,, \label{R1}
\end{equation}
for particles of zero rapidity.
The excess is due to the duration of emission, $\delta\tau$, of the
source in which particles with transverse momentum $p_\perp$ and
outward velocity $\beta_o=p_\perp/m_\perp$ travel a distance
$\beta_o\;\delta\tau$ towards the detector. Distances
perpendicular to this velocity  like, for example, $R_s$ are not
affected by the duration of emission and thus reflect the ``true'' transverse
size of the source, i.e., the region over which the source can be
considered homogeneous.
When flow is included the relation (\ref{R1}) is still valid in the analyses
of \cite{Heinz,Csorgo} whereas strong flow coupled to
surface emission can reduce $R_o$ more than $R_s$ \cite{Pratt} as well as
an ellipsoidal source \cite{CL,CV}.

Experimentally the HBT radius parameters, $R_{s,o,l,ol}$, are extracted from
measurements of the two-pion and two-kaon correlation function by
parametrizing it with the common gaussian form
\begin{equation}
  C(q_s,q_o,q_l)=1+\lambda\exp(-q_s^2R_s^2-q_o^2R_o^2-q_l^2R_l^2
  -2q_oq_lR_{ol}^2)\;.   \label{Cexp}
\end{equation}
Surprisingly, in relativistic heavy ion collisions the outward and
sideward radius parameters are measured to be similar
\cite{NA44,NA44QM,NA49,WA80,AGS} (and in a few cases the outward
size is even measured to be smaller than the sideward size
\cite{NA44,NA44QM} contradicting equation (\ref{R1})) within experimental
uncertainty.  According to equation (\ref{R1}) this implies that particles
freeze--out suddenly, $\delta\tau\ll R_i$, as in a ``flash'' \cite{CC},
in particular when resonance life-times are included \cite{CL,HH}.

 We want to point out that an {\it opaque} source emitting {\it away} from
its surface naturally leads to $R_o\ll R_s$ unless the duration of emission,
$\delta\tau$, is very long.  There is therefore a strong correlation
between the direction of the emitted particles and the emission zone -
contrary to cylindrically symmetric transparent sources without flow. 
In our model the measured sideward radius parameter samples the full 
source size, while the measured outward radius parameter
sample only a small surface region of the opaque source in the
direction of the emitted particles.  The opaque source is inspired by
hydrodynamical as well as cascade calculations of particle emission in
relativistic heavy ion collisions. In hydrodynamical calculations
particles freeze-out at a hypersurface that generally does not move
very much transversally until the very end of the
freeze--out \cite{hydro}. However, most hydrodynamical freeze--out
mechanisms as well as the analysis in \cite{Pratt}
do not include the directional condition that particles can
only be emitted away from the surface, though strong flow at the surface
has a similar effect. In cascade codes the last interaction points are
also found to be distributed in transverse direction around a mean
value that does not change much with time \cite{Humanic,RQMD,QGSM}.
 The thickness of the emission layer is
related to the particle mean free path $\lambda_{mfp}$, and is zero in
hydrodynamical calculations but several {\it fm}'s in cascade codes.
We shall in section 2 consider the parameters $R$ and 
$\lambda_{mfp}$ as constants and in section 3 
discuss effects of moving surfaces, temporally dependent
emission layers and transverse flow. 

\section{Surface emission and HBT Radius Parameters}

To quantify our statements we calculate the radius 
parameters $R_{s,o,l}$ for two simple
sources commonly employed to describe relativistic and
ultrarelativistic heavy ion collisions, namely a spherical and a
cylindrical symmetric but longitudinally expanding source. However, we
add the important requirement that the source should be opaque
emitting away from a surface layer of thickness $\sim\lambda_{mfp}$.

For the correlation function analysis of Bose-Einstein interference
from a source of size $R$ we consider two particles emitted a distance
$\sim$$R$ apart with relative momentum ${\bf q}=({\bf k}_1-{\bf k}_2)$
and average momentum, ${\bf K}=({\bf k}_1+{\bf k}_2)/2$. Typical heavy
ion sources in nuclear collisions are of size $R\sim5$ fm, so that
interference occurs predominantly when $q\sim\hbar/R\sim 40$
MeV/c. Since typical particle momenta are $k_i\simeq K\sim 300$ MeV,
the particles escape almost parallel (see Figure), i.e., $k_1\simeq
k_2\simeq K\gg q$.  The correlation function due to Bose-Einstein
interference of identical particles from an incoherent source is (see,
e.g., \cite{Heinz})
\begin{equation}
   C({\bf q},{\bf K})=1\;+\; |\frac{\int d^4x\;S(x,{\bf K})\;e^{iqx}}
   {\int d^4x\;S(x,{\bf K})}|^2 \,, \label{C}
\end{equation}
where $S(x,{\bf K})$ is a function describing the phase space density of the
emitting source.
With $qx={\bf q\cdot x}-{\bf q\cdot}$\mbox{\boldmath $\beta$}$_K\,t$,
where \mbox{\boldmath $\beta_K$}=${\bf K}/E_K$ is the pair velocity,
one can, by expanding to second order in $q_i\;R_i$ and compare to
equation (\ref{Cexp}), find the radius parameters 
$R_i$, {\it i=s,o,l,ol}, \cite{Heinz}
\begin{equation}
   R_i^2=\sigma(x_i-\beta_i\;t)  \,,\quad 
   R_{ol}^2=\langle(x_o-\beta_o\;t)\;(x_l-\beta_l\;t)\rangle \;, 
   \label{Ri}
\end{equation}
where the source average of a quantity ${\cal O}$ and its fluctuation
are defined by
\begin{equation}
  \langle{\cal O}\rangle\equiv
   \frac{\int d^4x\;S(x,K){\cal O}}{\int d^4x\;S(x,K)} \;,\quad
  \sigma({\cal O})\equiv \langle{\cal O}^2\rangle - \langle{\cal O}\rangle^2
. \label{O}
\end{equation} 
The reduction factor $\lambda$ in equation (\ref{Cexp}) may be
due to long lived resonances \cite{HH,CL}, coherence effects,
incorrect Gamov corrections \cite{BB} or other effects. It is found to
be $\lambda \sim 0.5$ for pions and $\lambda \sim 0.9$ for kaons.

\begin{figure}
\centerline{
\psfig{figure=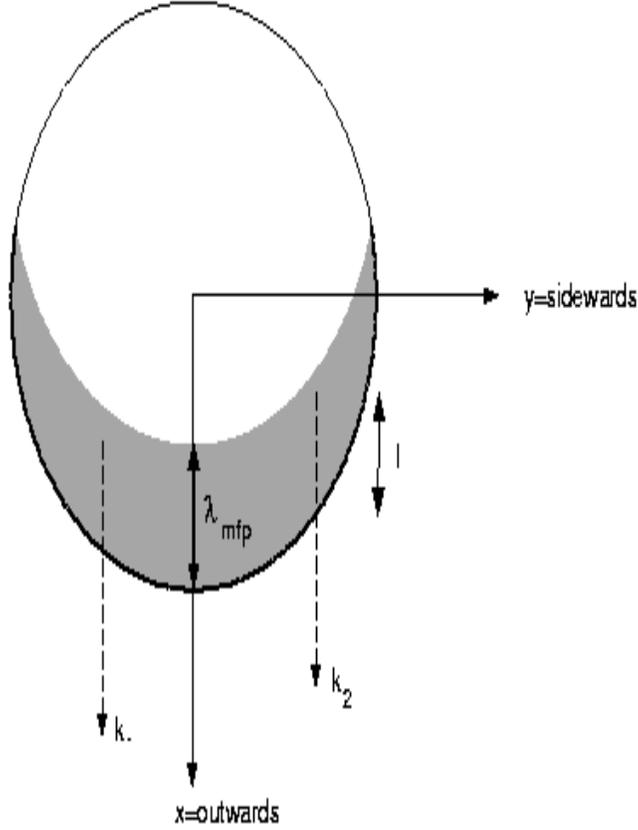,width=13cm,height=9.5cm,angle=-90}}
\caption{
Cross section of the interaction region perpendicular to the
longitudinal or z--direction. Particles have to penetrate a distance
$l$ out to the surface of the interaction region in order to escape and reach
the detector. The bulk part of the emitted particles comes from a surface
region of width $\lambda_{mfp}$, the mean free path of the particle.
}
\end{figure}

Generally a particle is emitted and escapes after it has made its
final interaction with any of the other particles in the source. If
the source is dense it is more
probable to emit from the surface than from the center.
In Glauber theory particles are emitted from a surface layer with a
depth of order of the mean free path, $\lambda_{mfp}$, with
probability 
\begin{equation}
   S(x,{\bf K}) \sim \exp(-\int_x^\infty dx'\sigma n(x'))
                = \exp(-l/\lambda_{mfp})  \,. 
\end{equation}
Here the integral runs over the particle trajectory from emission
point $x$ to the detector. Assuming an average density $n$ in the
emission layer, the distance $l$, that the particle has to pass
through is of order of the mean free path, $\lambda_{mfp}=1/\sigma n$
(see Fig. 1).  A similar surface layer was introduced by Grassi et
al. \cite{Grassi} in order to improve the standard Cooper-Frye
freeze-out in hydrodynamical calculations and to study the effect on
transverse momentum spectra.

In the following subsection we assume that the source size $R$ and
thickness of emission layer $\lambda_{mfp}$ are constants in time.
In section 3 will will generalize these results including
temporal dependences in connection with the final freeze-out of the source.

\subsection{Spherical Sources}

First we investigate a spherically symmetric source of
radius $R$. Including the temporal evolution of the source
emission function by  $S_t(t)$, the
phase space density of the source will have the approximate form
\begin{equation}
   S(x,K) \sim   \Theta(R^2-{\rm x^2-y^2-z^2})\;
          e^{-l/\lambda_{mfp}} \; S_t(t)    \label{SS} \,.
\end{equation}
For strict surface emission\footnote{
A spherical source similar to (\ref{SSd}) with strict surface
emission, $\lambda_{mfp}=0$, was applied to proton emission from
excited nuclei \cite{KP}. As proton decay times are long only the
finite duration of emission contribution was considered (i.e, the last term
in equation (\ref{Ros}).) }
($\lambda_{mfp}=0$) the source reduces to
\begin{equation}
   S(x,K) \sim   \cos\theta \;\Theta(\cos\theta)
                 \delta(R^2-{\rm x^2-y^2-z^2})\;
                 \; S_t(t) \,. \label{SSd}
\end{equation}
where $\theta$ is the polar angle with respect to the outward direction
along ${\bf K}$; we will in the following orient our cartesian axes such
that the {\rm x}-axis is in the outward direction.
The geometric factor $\cos\theta={\rm x}/R$ suppresses the peripheral zones and
the $\Theta(\cos\theta)=\Theta({\rm x})$ factor insures that
particles are only emitted {\it away} from the surface, i.e., only
particles from the surface layer of the half hemisphere directed
towards the detector will reach it whereas particles from the other
hemisphere will interact on their passage through the
source.
 
The temporal emission is determined by $S_t(t)$. It is commonly
approximated by a gaussian, $S_t(t)\sim\exp(-(t-t_0)^2/2\delta t^2)$,
around the source mean life-time, $\tau_0$ with width or fluctuation,
$\delta\tau$, which is the duration of emission.  These gaussian
parameters approximate the average emission time, $\langle t\rangle$
and the variance or fluctuation, $\sigma(t)$, for a general source,
respectively.

Orienting 
our coordinate system with outward or {\rm x}-axis along ${\bf K}$
we have \mbox{\boldmath $\beta_K$}=${\bf K}/E_K-(\beta_o,0,0)$ (see Figure 1)
when the pair rapidity has been boosted longitudinally to their
center-of-mass system, $Y=0$. 
In that case $l=(\sqrt{R^2-{\rm y^2-z^2}}-{\rm x})$.
When $\lambda_{mfp}\ll R$ we obtain from equations (\ref{C}-\ref{O})
\begin{eqnarray}
   R_o^2 &=& \frac{1}{18}R^2 \;+\;\frac{29}{36}\lambda_{mfp}^2
          \;+\;\beta_o^2\sigma(t) \,, \label{Ros}\\
   R_s^2 &=&\frac{1}{4} R^2 \;-\; \frac{1}{8}\lambda_{mfp}^2 \,, \label{Rss}\\
   R_l^2 &=&  R_s^2 \; ,          \label{Rls}
\end{eqnarray}   
and $R_{ol}$ vanishes. Aside from the $\beta_o^2\;\sigma(t)$ term, we
notice that an opaque ($\lambda_{mfp}\ll R$) source has
$R_o^2=2R_s^2/9$, i.e., the outward radius parameter seems much 
smaller than the
actual source size as well as $R_s$.  This is a simple geometrical
effect that arises because no particles are observed from the dark
side of the source (see Figure 1) leading to an expectation value of
$\langle {\rm x}\rangle$ that is displaced from the center and only
slightly smaller than $\sqrt{\langle {\rm x}^2\rangle}$. For
comparison, a transparent ($\lambda_{mfp}=\infty$) and spherically
symmetric source has the same extent in all directions and one finds
$R_s=R_o=R_l=R/\sqrt{5}$. Including the term due to the duration of
emission, $\beta_o^2\;\sigma(t)$, the outward HBT radius parameter is
necessarily larger than the sideward.

\subsection{Cylindrical Sources}

At very high energies longitudinal expansion is important and we
employ a locally thermal but longitudinally expanding 
(with Bjorken flow $u={\rm z}/t$) source as in \cite{Heinz,Csorgo}
\begin{equation}
   S(x,K) \sim \Theta(R^2-{\rm x^2-y}^2)
             \exp\left( -\frac{m_\perp\cosh(Y-\eta)}{T}
                        -\frac{l}{\lambda_{mfp}} \right)\; S_\tau(\tau)\,.
\end{equation}
Here, $\tau=\sqrt{t^2-{\rm z}^2}$ is the invariant or proper
time and $\eta=0.5\ln(t+{\rm z})/(t-{\rm z})$ the space-time rapidity.
$R$ is the transverse source size, $m_\perp$ the 
transverse mass and $Y$ the rapidity of the particles.
The distance that the particles have to pass through matter in the 
transverse direction is
 $l \simeq (\sqrt{R^2-{\rm y}^2}-{\rm x})$.
Actually, when $\eta\ne Y$ the particles pass
a distance longitudinally, which will lead to a minor reduction of $R_l$,
but as we concentrate on the transverse radius parameters
we will ignore this effect in the following.
 From equations (\ref{C}-\ref{O}) we obtain when $\lambda_{mfp}\ll R$
\begin{eqnarray}
   R_o^2 &=& \left(\frac{2}{3}-(\frac{\pi}{4})^2\right)R^2
   \;+\;\left(\frac{7}{6}-\frac{\pi^2}{32}\right)\lambda_{mfp}^2 
   \;+\;    \beta_o^2\; \sigma(\tau)
      \label{Ro} \,, \\
   R_s^2 &=& \frac{1}{3}R^2\;-\;\frac{1}{6}\lambda_{mfp}^2 \,, \label{Rs}\\
   R_l^2 &\simeq& \langle\tau^2\rangle \;\frac{T}{m_\perp} 
		\label{Rl} \,,
\end{eqnarray}  
for $Y=0$ (which leads to $R_{ol}=0$) and keeping only leading orders
in $\sigma(\tau)$ and $T/m_\perp$ (see, e.g., \cite{Heinz} for higher
orders).  Again we notice that for a short--lived ($\beta_o\;\sigma(\tau)^{1/2}
\ll R$) and opaque ($\lambda_{mfp}\ll R$) source the outward radius parameter
seems much smaller than the actual source size as well as the sideward
radius parameter, $R_o=0.22R=0.39R_s$ (see Figure 2). For a short--lived
transparent source ($\lambda_{mfp}=\infty$) the transverse 
radius parameters are
the same $R_s=R_o=R/2$.  The experimental measurements find, that
$R_o\sim R_s$. This implies a very long duration of emission,
$\beta_o\;\sigma(\tau)^{1/2}\sim R_s$, for an opaque source and is in sharp
contrast to the very short duration of emission the same data would
imply for a transparent source.

\section{Temporal dependence and Freeze--out}

Eventually all particles created in the high energy nuclear collisions
must freeze--out. In cascade codes the emission changes
from strict surface emission shortly after the collision starts
to an emission layer, which increases in
thickness to the size of the system. At this stage volume emission takes 
over and, aside from long lived resonance decays, the source freezes out 
\cite{Humanic,RQMD,QGSM}.
This picture is somewhat different in hydrodynamic codes, where the
emission is strictly from the surface at all times, but the surface 
eventually moves inward and the source freeze out when the surface
hits the center \cite{hydro}. However, just before freeze--out the hydrodynamic
hypersurface moves inward with superluminous speed. Again this stage
is better characterized by volume freeze--out rather than by surface emission.
The hydrodynamic results may become more similar to those of the cascade
models, if we include emission from a surface layer of finite thickness,
as in \cite{Grassi}, as well as a cascade among the hadronized particles.

\begin{figure}
\centerline{
\psfig{figure=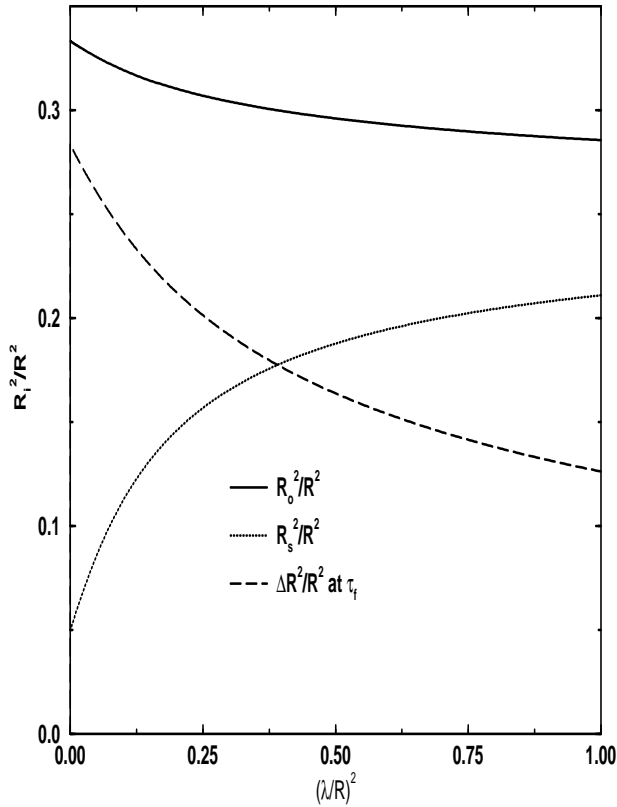,width=13cm,height=9.5cm,angle=-90}}
\caption{
Sideward and outward HBT radius parameters as function 
of the relative thickness of surface layer thickness to source size. 
The small mean free path limits are given in equations (\protect{\ref{Ros}}) 
and (\protect{\ref{Rss}}) and for
$\lambda_{mfp}\gg R$ they approach $R_s=R_o=R/2$. The duration of emission
contribution $\beta_o^2\;\sigma(\tau)$ is not included in
$R_o^2$. The strength of the opacity effect, $\Delta R^2/R^2$, 
is estimated from equation (\protect{\ref{Ropaque}}) at 
the freeze--out time $\tau_f$.}
\end{figure}

The freeze-out can be described by allowing the layer of emission or
$\lambda_{mfp}$ to increase with time after the collision. According
to the Bjorken scaling model the mean free path increases linearly with proper
time $\tau$
\begin{eqnarray}
  \lambda_{mfp}(\tau) = \frac{1}{\bar{\sigma} n(\tau)} =
            \frac{\pi R^2}{\bar{\sigma} dN/dy} \tau
          \equiv \lambda_{mfp}(\tau_f)\frac{\tau}{\tau_f} \, ,\label{lambda}
\end{eqnarray}
since the particle density drops inversely with proper time due to the
one-dimensional expansion in longitudinal direction. In cascade codes
the size of the emission layer also increases with time up to freeze-out.
Assuming such a
linear increase a simple estimate of the HBT radius parameters can be
made.  From the definition in equation (\ref{Ri}) we find, that the
HBT radius parameters are the time-averaged values of the
$\lambda_{mfp}$ dependent HBT radius parameters
\begin{eqnarray}
   R_i^2 = \frac{2}{\tau_f^2} \int_0^{\tau_f} 
           R_i^2(\lambda_{mfp}(\tau))\;\tau\;d\tau
         = \int_0^{\lambda_{mfp}^2(\tau_f)}  R_i^2(\lambda_{mfp}(\tau))
           \;\frac{d\lambda_{mfp}^2}{\lambda_{mfp}^2(\tau_f)}  \,.
\end{eqnarray}
Thus the HBT radius parameters are the average values of the mean free path
dependent radius parameters of Figure (2) up to $\lambda_{mfp}^2(\tau_f)$.  
If we include now the duration of emission the difference is
\begin{eqnarray}
   R_o^2-R_s^2 &=& \int_0^{\lambda_{mfp}^2(\tau_f)}  
           \left( R_o^2(\lambda_{mfp})-R_s^2(\lambda_{mfp})\right)
           \;\frac{d\lambda_{mfp}^2}{\lambda_{mfp}^2(\tau_f)} 
                 +\beta_o^2\;\sigma(\tau) \nonumber\\
          &\equiv& -\Delta R^2 +\beta_o^2\;\sigma(\tau) 
           \,. \label{Ropaque}
\end{eqnarray}
Here $\Delta R^2$ is the average distance between $R_s^2$
and $R_o^2$ in Figure 2, i.e., the area between the curves for $R_s^2$
and $R_o^2$ up to $\lambda_{mfp}^2(\tau_f)$ and divided by that same
factor $\lambda_{mfp}^2(\tau_f)$.  The opacity difference $\Delta
R^2$ is plotted as dashed line in Figure (2). Though it
decreases with increasing $\lambda_{mfp}(\tau_f)$, it is a substantial
fraction of $R_s^2$ for mean free paths at freeze-out of order the
size of the system.

It is instructive to consider the recent and relevant example of
central $Pb+Pb$ collisions at energy 160 A$\cdot$GeV.  The NA44 data
finds for the pion HBT radius parameters $R_s\simeq R_o\sim 4.5-5.0$
fm and $R_l\simeq5-6$ fm \cite{NA44}.  The average transverse momentum
was $p_\perp\simeq 165$ MeV such that $\beta_o=p_\perp/m_\perp\simeq
0.76$. From the transverse momentum slopes of pions, kaon, protons and
deuterium in \cite{NA44slopes} one finds a temperature $T\sim 120$ MeV
and transverse flow $v\sim 0.6c$. Larger transverse flow and smaller
temperatures can, however, not be excluded.  The transverse flow
effect on the HBT radius parameters is small for these $p_\perp$ and
$v$ values \cite{HBTsub}.  If we assume that the source emits pions at
a constant rate per surface element, we can estimate from the
longitudinal HBT radius parameter, equation (\ref{Rl}), the freeze-out
time $\tau_f=(2\;m_\perp/T)^{1/2}\;R_l\sim 11$ fm/c.  As seen from
equations (\ref{Rs}) the sideward radius parameter is relatively less
affected by the $\lambda_{mfp}^2$ term and we extract an initial
transverse source size $R\sim 3^{1/2}\;R_s\sim 9$ fm from the NA44
sideward HBT radius parameter.

To determine the outward HBT radius parameter we have to estimate the
mean free path, for example, from equation (\ref{lambda}).  In central
$Pb+Pb$ collisions at 160 A$\cdot$GeV the particle rapidity density is
$dN/dy\sim 600$ around midrapidity. If we take for the pion cross
section $\bar{\sigma}_\pi\sim 20$ mb, the resulting mean free path is
very small, $\lambda_{mfp}(\tau_f)\sim1$ fm. However, near the surface
the density will be lower than the average density used in this
estimate and we would rather expect $\lambda_{mfp}(\tau_f)\sim R$
since pion freeze-out occur when their mean free path is of order the
size of the system.  For example, $\Delta R^2\simeq 0.13
\;(0.07)\;R^2$ for $\lambda_{mfp}=1.0\;(2.0)\;R$ respectively and the
opacity effect leads to a corresponding difference $\Delta
R^2\simeq 16\;(9)$ fm$^2$.  Equation (\ref{Ropaque}) clearly
demonstrates the significance of the opacity effect in reducing the
outward radius parameter with respect to the sideward.  The duration
of emission compensates this difference to some extent.  Assuming
constant particle emission per surface element up to freeze-out, we
find $\beta_o^2\;\sigma(\tau)=\beta_o^2\;\tau_f^2/18 \simeq 4$ fm$^2$.
This contribution due to the duration of emission is thus smaller than
the reduction due to the opacity contribution and consequently one
would expect $R_o<R_s$. However, a number of other fluctuations such
as moving surfaces, short lived resonances, and other effects also add
to $R_o^2$ \cite{HBTsub}.

We notice that the transverse source size is only a few $fm$ larger
than the geometrical size of $Pb$ $\sim7$ fm.  Some small expansion of
the source seems to take place before final freeze-out. The average
emission time is $\sqrt{\langle\tau^2\rangle}=\tau_f/\sqrt{2}\simeq 8$
fm/c.  This result is entirely consistent with the freeze-out time
needed to explain the enhancement in the $\pi^-/\pi^+$ ratio at low
$p_\perp$ in the same $Pb+Pb$ collisions due to Coulomb repulsion
\cite{Barz}.

\section{Summary and Outlook}

We have addressed the significant difference in HBT radius parameters
between opaque and transparent sources and have calculated the radius
parameters and their dependence on the mean free path for simple
spherical and longitudinal expanding models. A short--lived and very
opaque source or ``black body'' has much smaller outward than sideward
radius parameter.  However, a long duration of emission as well as
other fluctuations can compensate the reduction due to the opacity
effect leading to comparable $R_o\simeq R_s$ as measured in several
experiments.  In the case of central $Pb+Pb$ collisions at 160
A$\cdot$GeV the opacity reduction was estimated to be a significant
fraction of the outward HBT radius parameter and larger than the
contribution from a source emitting during all its life-time.  Other
fluctuations like moving surfaces, short lived resonances, and other
effects do, however, also add to $R_o^2$ and we expect that this is
the reason that the sideward and outward HBT radius parameters are
measured to be very similar in relativistic heavy ion collisions.  We
find that pions do not appear as in a ``flash''. In fact a long
duration of emission of order the life-time of the source is possible
and consistent with an outward HBT radius parameter smaller or
comparable to the sideward due to the opacity effect.  In contrast, a
simple transparent source would necessarily have a very short duration
of emission as implied by equation (1).

The Cooper-Frye freeze-out condition in hydrodynamic models does not
take the opacity effect into account and generally one finds considerably
larger outward than sideward source radius parameters due to the long
freeze--out time \cite{hydro}. Cascade codes have implicitly opacities
build in through rescatterings and do find a directional effect
\cite{Humanic}, but also long duration of emission and mean free paths
\cite{RQMD} are found leading to larger outward radius parameters than
sideward.

Other effects like, e.g., transverse flow, can also reduce the outward
HBT radius parameter more than the sideward.  At larger transverse
momenta the sideward and outward HBT radius parameters are reduced by
transverse flow and the longitudinal scales by the factor $1/m_\perp$.
This may explain the decrease of all the HBT radius parameters with
increasing $p_\perp$ as found in the NA44 experiments \cite{NA44}.

By studying the rapidity and $p_\perp$ dependence of the HBT radius
parameters the duration of emission contribution,
$\beta_o^2\;\sigma(\tau)$, may be separated from the opacity effect,
$\Delta R^2$, as well as other fluctuations.  However, this may be
non-trivial if the mean free path, duration of emission and resonance
contributions are rapidity and $p_\perp$ dependent. Transverse flow
may also add a strong $p_\perp$ dependence of the radius parameters
but the magnitude of transverse flow can approximately be determined
from transverse momentum spectra \cite{NA44slopes}.  HBT radius
parameters and transverse flow increase for heavier ions
colliding. The nuclear $A$-dependence can therefore provide additional
information on freeze-out times, sizes, transverse flow and opacity of
the sources.

\section*{Acknowledgements}
We would like to thank Larry McLerran and Scott Pratt
for stimulating discussions.

\end{document}